\documentclass{PoS}

\def\beq{\begin{equation}}
\def\eeq{\end{equation}}
\def\bea{\begin{eqnarray}}
\def\eea{\end{eqnarray}}
\def\D0{D\O }

\title{Selected updates on semileptonic $B$ decays and $|V_{xb}|$ determination}

\ShortTitle{Semileptonic $B$ decays and $|V_{xb}|$ determination}

\author{\speaker{Giulia Ricciardi}
\\
        Dipartimento di Fisica E. Pancini, Universit\`a  di Napoli Federico II,
and
INFN, Sezione di Napoli,
Complesso Universitario di Monte Sant'Angelo, Via Cintia,
I-80126 Napoli, Italy\\
        E-mail: \email{giulia.ricciardi@na.infn.it}}

\abstract{We summarize  selected up-to-date results related to semileptonic $B$ meson decays, namely results on the exclusive determination of the parameter $|V_{cb}|$ of the CKM matrix, on the inclusive determination of  $|V_{ub}|$,  and on $R(D^{(*)})$.
}

\FullConference{XIII Quark Confinement and the Hadron Spectrum - Confinement2018\\
		31 July - 6 August 2018\\
		Maynooth University, Ireland}

\begin{document}

\section{Introduction}
\label{intro}
The parameters  $|V_{xb}|$  of the Cabibbo-Kobayashi-Maskawa (CKM) matrix
play a central role in the analyses of the unitarity triangle and in testing the Standard Model (SM). The long-standing tension among their values, depending on whether they  are  extracted  using exclusive or inclusive semi-leptonic $B$ decays, also  motivates  more and more precise theoretical and experimental investigations~\footnote{For brief overviews see for example \cite{Ricciardi:2013xaa,Ricciardi:2014aya,Ricciardi:2014iga,Ricciardi:2016pmh, Ricciardi:2017lne,  Koppenburg:2017mad} and references therein.}. We briefly  discuss recent results on the exclusive determination of $|V_{cb}|$, on the inclusive determination of  $|V_{ub}|$,  and on $R(D^{(*)})$.

 \section{Exclusive $|V_{cb}|$ determination}
\label{subsectionExclusive decays}

We can express the  differential ratios for the semi-leptonic CKM favoured decays $B \to D^{(\ast)} \ell \nu$
 in terms of the  recoil parameter $\omega = p_B \cdot p_{D^{(\ast)}}/m_B \, m_{D^{(\ast)}}$, which corresponds to the energy transferred to the leptonic pair. For negligible lepton masses ($\ell=e, \mu)$, each ratio depends
on one form factor, ${\cal F}(\omega)$  for $B \to D^{\ast} l  \nu$ and   ${\cal G}(\omega)$ for $B \to D l  \nu$,  and the phase space vanishes at the  no-recoil point $\omega=1$ in both cases. Summarizing, we can write
\begin{eqnarray}
&\frac{d\Gamma}{d \omega}(B \rightarrow D^\ast\,\ell {\nu})&
\propto G_F^2    (\omega^2-1)^{\frac{1}{2}}
 |V_{cb}|^2   {\cal F}(\omega)^2
 \nonumber \\
 &\frac{d\Gamma}{d \omega} (B \rightarrow D\,\ell {\nu})&  \propto
G_F^2\,
(\omega^2-1)^{\frac{3}{2}}\,
 |V_{cb}|^2   {\cal G}(\omega)^2
 \label{diffrat}
\end{eqnarray}
%
 In the
heavy quark limit,
both form factors are related to a single Isgur-Wise
function,  ${\cal F(\omega) }= {\cal G(\omega) } = {\cal  \xi (\omega) }  $, which is
normalized to unity at zero recoil,  that is  ${\cal \xi (\omega}=1) =1 $.
There are
non-perturbative corrections  to this prediction, expressed at the zero-recoil point by the heavy quark symmetry  under the form of
powers of $\Lambda_{QCD}/m$, where $m= m_c$ and $m_b$. Other corrections are perturbatively calculable radiative
corrections from hard gluons and photons.

In order to extract the CKM factors, we  need  not only to compute the form factors, but also to measure experimental decay rates, which
  vanish at zero-recoil.
Therefore,  experimental points are extrapolated to zero recoil, using a parametrization of the
dependence on $\omega$ of the form factor.
Several recent extrapolations to zero recoil adopt a parametrization where
 $\omega$ is mapped onto a complex variable $z$ via the conformal transformation
$
z= (\sqrt{\omega+1}-\sqrt{2})/(\sqrt{\omega+1}+\sqrt{2})
$. The form factors may be written in form of an
expansion in $z$, which converges rapidly in the kinematical
region of heavy hadron decays. The coefficients of the expansions  are subject to unitarity  bounds  based  on  analyticity.
 Common examples  are the
 CLN
(Caprini-Lellouch-Neubert) \cite{Caprini:1997mu},
 the BGL
(Boyd-Grinstein-Lebed)  \cite{Boyd:1994tt} and the  BCL (Bourrely-Caprini-Lellouch) \cite{Bourrely:2008za} parameterizations.
They are  all  constructed  to  satisfy  the  unitarity  bounds, but the CLN approach differs
mostly  in its reliance  on  next-to-leading  order
HQET  relations  between  the  form  factors.
Because of such additional constraints, the accuracy of the CLN approach has been questioned in both $B \to D\, l \nu  $ \cite{Bigi:2016mdz} and $B \to D^\ast \, l \nu$ \cite{Bigi:2017njr, Grinstein:2017nlq} channels.
The way out of parametrization dependence is to compute the
form factors at nonzero
recoil values. In the case of $B \to D\ell \nu$ decays, the form factors in the unquenched lattice-QCD approximation
are available for a range of non-zero recoil momenta since 2015.
Lastly, we remark that combined fits to $B \to D^{(*)}\ell \nu$
differential  rates  and  angular  distributions can help constraining the form factors and
allow not only the extraction of $|V_{cb}| $ but also of form factor ratios under various  scenarios \cite{Bernlochner:2017jka}.

\subsection{$B \to D^{\ast} \ell  \nu$ channel}

The form factor ${\cal F}(1)$ at zero recoil for the $B \to D^{\ast} \ell  \nu$ channel,
in the lattice unquenched $N_f= 2+1$  approximation, has been estimated first by the FNAL/MILC collaboration, which used Wilson fermions for both $c$ and $b$
heavy quarks \cite{Bailey:2014tva}
\beq  {\cal F}(1)
=0.906\pm 0.004 \pm  0.012  \label{VcbexpF2}  \eeq
The first error is statistical and the second one  is the sum in quadrature of all systematic errors.
The total uncertainty is around  the (1-2)\% level.
The largest error is the heavy quark discretization error related to the Fermilab action. The  LANL/SWME collaboration is working to reduce it by using an improved version of the Fermilab action, the Oktay-Kronfeld action \cite{Bailey:2017xjk}. Other preliminary results concern analyses at non-zero recoil \cite{Aviles-Casco:2017nge} and with M$\mathrm{\ddot{o}}$bius domain-wall quarks,  at zero and non-zero recoil, from $N_f= 2+1$
QCD  \cite{Kaneko:2018mcr}.
A  recent value of the form factor ${\cal F}(1)$ has been presented by the HPQCD collaboration, which used the fully relativistic
 HISQ (Highly
improved
staggered quarks) action for light, strange and charm quarks and the  NRQCD (Nonrelativistic QCD) action for the bottom quark \cite{Harrison:2017fmw}
\beq  {\cal F}(1)
=0.895\pm 0.010 \pm  0.024  \label{VcbexpF2HISQ}  \eeq
Both the  results in Eq. (\ref{VcbexpF2}) and Eq. (\ref{VcbexpF2HISQ}) are in good agreement.
%
%
%
The form factor in Eq. (\ref{VcbexpF2}) has been used by the latest $|V_{cb}|$ determinations from the Heavy Flavour and Lattice Averaging Groups, HFLAV and FLAG respectively.
 Using  the CLN parametrization,
the 2016 HFLAV average \cite{Amhis:2016xyh}
gives
\beq |V_{cb}| = (39.05 \pm 0.47_{\mathrm{exp}} \pm 0.58_{\mathrm{th}} ) \, \mathrm{x} \, 10^{-3} \label{ll1} \eeq
where the first uncertainty is experimental and the second error is theoretical (lattice QCD
calculation and electro-weak correction).
The 2016 FLAG  $N_{f}=2+1$ $|V_{cb}|$  average value yields \cite{Aoki:2016frl}
\beq |V_{cb}| = (39.27 \pm 0.49_{\mathrm{exp}} \pm 0.56_{\mathrm{latt}}) \times\, 10^{-3} \label{ll1flag} \eeq
This average employs  the 2014 HFLAV experimental average  \cite{Amhis:2014hma} ${\cal F}(1) \eta_{\mathrm{EW}} |V_{cb}|= (35.81 \pm 0.45) \times\, 10^{-3}$ and  the value $\eta_{\mathrm{EW}} = 1.0066$ \cite{Sirlin:1981ie}.


In 2017,  for the first time, the unfolded
fully-differential decay rate and associated covariance matrix have been published, by the Belle collaboration  \cite{Abdesselam:2017kjf}, prompting several independent determinations of the value of $|V_{cb}| $ under different theoretical approaches \cite{Grinstein:2017nlq,Bigi:2017njr, Bigi:2017jbd, Jaiswal:2017rve, Bernlochner:2017xyx, Harrison:2017fmw}.
Most notably, using the Belle measurement  \cite{Abdesselam:2017kjf}, it has been argued
that previous determinations using the CLN form factor parameterisation could be affected by underestimated uncertainties,  and  therefore  form  factor  analyses  based
on BGL  should be preferred
\cite{Grinstein:2017nlq,Bigi:2017njr, Bigi:2017jbd}.
The previous analyses, in the BGL framework,
appear consistent with each other and give,  using t lattice  zero recoil form factors, the values
 \bea
 |V_{cb}| &=& (41.9^{+2.0}_{-1.9}) \, \times \, 10^{-3}  \quad \cite{Grinstein:2017nlq} \nonumber \\
 |V_{cb}| &=& (41.7^{+2.0}_{-2.1}) \, \times\, 10^{-3} \quad \cite{Bigi:2017njr}
 \eea
The central values are higher than the corresponding values in the CLN parametrization, and closer to the values from the
inclusive  approach.
On the other side, it has also been argued that
 fits that yield the higher values of
$|V_{cb}|$
 suggest large violations of heavy quark symmetry
and   tension  with   lattice  predictions of  the  form factor ratios \cite{Bernlochner:2017xyx}.

This year for the first time the Belle Collaboration has performed fits with both the CLN and BGL form factor parameterizations, obtaining \cite{Abdesselam:2018nnh}
\bea
{\cal F}(1) \eta_{\mathrm{EW}} |V_{cb}| &=& (35.06 \pm 0.15 \pm 0.54) \times\, 10^{-3} \qquad \mathrm{CLN \; \; fit} \nonumber \\
{\cal F}(1) \eta_{\mathrm{EW}} |V_{cb}| &=& (38.73 \pm 0.25 \pm 0.60) \times\, 10^{-3} \qquad \mathrm{BGL \; \; fit}
\eea
The first uncertainty is statistical, the second one is systematic.
This is the experimentally    most  precise  determination
performed  with  exclusive  semileptonic
B
decays. Taking the value for the form factor in Eq. (\ref{VcbexpF2}) and $\eta_{\mathrm{EW}}=1.006$ \cite{Sirlin:1981ie}, they find \cite{Abdesselam:2018nnh}
\bea
 |V_{cb}| &=& (38.4 \pm 0.2 \pm 0.6 \pm 0.6) \times\, 10^{-3} \qquad \mathrm{CLN \; \; +\; \; LQCD } \nonumber \\
 |V_{cb}| &=& (42.5 \pm 0.3 \pm 0.7 \pm 0.6) \times\, 10^{-3} \qquad \mathrm{BGL \; \;+\; \; LQCD}
\eea
The value of
branching fraction is found to be insensitive to the choice of the parameterization.  Both sets of fits give acceptable $\chi^2/$ndf:
therefore the data do not discriminate between the parameterizations.

Form factor estimates via zero recoil sum rules
  \cite{Faller:2008tr,Gambino:2010bp, Gambino:2012rd} give, in general, relatively higher values of $|V_{cb}|$ and
a theoretical error  more than twice the error in the lattice determinations.  Recently,
 form factors determinations from $B$ meson light-cone sum rules beyond leading twist have been  presented \cite{Gubernari:2018wyi}.

%

\subsection{The $B \to D \ell  \nu$ channel}

For $ B \rightarrow D \, \ell \, \nu$ decay, form factors are available for a range of recoil momenta  by both the FNAL/MILC collaboration and the  HPQCD Collaboration
 \cite{Lattice:2015rga, Na:2015kha}.
 By parameterizing the dependence on momentum transfer
using  the BGL parameterization, the former collaboration has determined
$ |V_{cb}| $
from the relative normalization over the entire range of recoil momenta
 \cite{Lattice:2015rga}
\beq
 |V_{cb}| =(39.6 \pm 1.7_{\mathrm{exp+QCD}} \pm 0.2_{\mathrm{QED}})   \, \mathrm{x} \, 10^{-3}
\eeq
The   average value is almost the same than the  one inferred from
$ B \rightarrow D^\ast \, \ell \, \nu$ decay by the same FNAL/MILC collaboration, see Eq. (\ref{ll1}).
The HPQCD Collaboration  has performed
a  joint  fit to  lattice  and 2009  BaBar  experimental  data  \cite{Aubert:2009ac} and extracted
the CKM matrix element  $|V_{cb}| $ using the CLN parametrization \cite{Na:2015kha}
\beq
|V_{cb}|= (40.2 \pm 1.7_{\mathrm{latt+stat}} \pm 1.3_{\mathrm{syst}})   \, \mathrm{x} \, 10^{-3}
\eeq
The  first  error  consists  of  the  lattice  simulation  errors  and  the  experimental  statistical  error  and
the  second  error  is  the  experimental  systematic  error.

%

In  2015 the decay $ B \rightarrow D\, \ell \, \nu$ has also been measured in fully reconstructed events  by the Belle collaboration \cite{Glattauer:2015teq}. They have  performed a fit to the CLN parametrization, which  has been used to determine $\eta_{EW} {\cal G}(1) |V_{cb}|$. By using the form-factor normalization ${\cal G}(1)$ found by the FNAL/MILC Collaboration
\cite{Lattice:2015rga}, and
 $\eta_{EW}   \simeq 1.0066 $ \cite{Sirlin:1981ie}, they obtain \cite{Glattauer:2015teq}\beq
 |V_{cb}|=(39.86 \pm 1.33) \times 10^{-3}\eeq The Belle Collaboration has also obtained
a slightly more precise
result (2.8\% vs.  3.3\%)  by
 exploiting  lattice data at non-zero recoil and  performing   a combined fit to the BGL form factor. It yields \cite{Glattauer:2015teq}
\beq
 |V_{cb}|=(40.83 \pm 1.13) \times 10^{-3}
\eeq

Global fit results in the BGL parametrization   are in agreement with the previous determinations, giving the value \cite{Bigi:2016mdz}
\beq
 |V_{cb}|=(40.49 \pm 0.97) \times 10^{-3} \eeq
Here the lattice results \cite{Lattice:2015rga, Na:2015kha}, as well as  Belle \cite{Glattauer:2015teq} and Babar \cite{Aubert:2009ac} data, have been used.

\section{Inclusive $|V_{ub}|$ determination}

In order to extract  $|V_{ub}|$  from semileptonic $B \to X_u \ell \nu$ decays one has to  reduce the $b \to c$ semileptonic background through experimental  cuts. Such cuts enhance the relevance of the so-called threshold region in the phase space, jeopardizing the use of operator product expansion techniques. In order to face this problem, that is absent in the inclusive determination of $|V_{cb}|$,   different theoretical schemes have been devised, which are  tailored
to analyze data in the threshold region,  but  differ
in their treatment of perturbative corrections and the
parametrization of non-perturbative effects.
In  Table \ref{phidectab04} we present the results for four theoretical different approaches, as analyzed
by BaBar \cite{Lees:2011fv, Beleno:2013jla}, Belle \cite{Urquijo:2009tp}  and  HFLAV  \cite{Amhis:2016xyh} collaborations, that is: ADFR  \cite{Aglietti:2004fz, Aglietti:2006yb,  Aglietti:2007ik}, BLNP
 \cite{Lange:2005yw, Bosch:2004th, Bosch:2004cb}, DGE\cite{Andersen:2005mj} and  GGOU  \cite{Gambino:2007rp} \footnote{Artificial neural networks have also been used to parameterize the shape functions and  extract $|V_{ub}|$ in the GGOU framework \cite{Gambino:2016fdy}. The results are in good agreement with the original paper.}.
Although conceptually quite different, all these approaches
lead to roughly consistent results when the same inputs are used and the
theoretical errors are taken into account.
%
\begin{table}[h]
\centering
\caption{Status of inclusive $|V_{ub}|$  determinations.}
\label{phidectab04}
\begin{tabular}{lrrrr}
 \hline
 { \color{red}{ Inclusive decays}} &
& {\color{red}{ $ |V_{ub}| \times  10^{3}$}}
  \\
\hline
& { \color{blue}{ ADFR }}  \cite{Aglietti:2004fz, Aglietti:2006yb,  Aglietti:2007ik}  &  { \color{blue}{  BNLP  }}   \cite{Lange:2005yw, Bosch:2004th, Bosch:2004cb}&   { \color{blue}{  DGE  }}   \cite{Andersen:2005mj} &  { \color{blue}{   GGOU  }}    \cite{Gambino:2007rp} \\
\hline
HFLAV 2016 \cite{Amhis:2016xyh} & $4.08 \pm 0.13^{+ 0.18}_{-0.12}$ & $ 4.44 \pm 0.15^{+0.21}_{-0.22}  $  & $4.52 \pm 0.16^{+ 0.15}_{- 0.16}$ &
$4.52 \pm  0.15^{ + 0.11}_ { -0.14} $  \\
BaBar 2011  \cite{Lees:2011fv} &  $4.29 \pm 0.24^{+0.18}_{-0.19}  $  & $4.28 \pm 0.24^{+0.18}_{-0.20}  $    & $4.40 \pm 0.24^{+0.12}_{-0.13}  $
& $4.35 \pm 0.24^{+0.09}_{-0.10}  $ \\
 Belle 2009 \cite{Urquijo:2009tp} & $4.48 \pm 0.30^{+0.19}_{-0.19}  $ & $ 4.47 \pm 0.27^{+0.19}_{-0.21}  $ &  $4.60 \pm 0.27^{+0.11}_{-0.13}  $ & $4.54 \pm 0.27^{+0.10}_{-0.11}  $ \\
\hline
\end{tabular}
\end{table}
Generally, these values are  higher compared to corresponding exclusive determinations. The tension is of the order of $2-3\sigma$, depending on the chosen framework.
When averaged, the ADFR value is lower than the ones obtained with the other three approaches, and closer to the exclusive values.
By taking the arithmetic average of the
results obtained from these  four different QCD predictions of the partial rate, the Babar collaboration gives \cite{Lees:2011fv}
\beq
|V_{ub}|=(4.33 \pm 0.24_{\mathrm{exp}} \pm 0.15_{\mathrm{th}}) \times 10^{-3}
\label{VinclBabar}
\eeq.
%

The latest analyses and theoretical progresses in semi-leptonic heavy to light decays involve power corrections \cite{Gambino:2016jkc,Gunawardana:2017zix}, global fits  to
inclusive rare and semileptonic data \cite{Gambino:2016fdy, Bernlochner:2011di} and, most recently,
exploratory lattice calculations of inclusive B meson semileptonic decay \cite{Hashimoto:2018gld}.
At Belle II prospects are good for improvements of the
$|V_{ub}|$ determination, with both inclusive and exclusive approaches,
thanks to more data and better reconstruction performance \cite{DeNardo:2018glk}.


\section{Exclusive $b \to c$ decays into heavy leptons}

In the SM the  couplings to the $W^\pm$ bosons are universal for all leptons and processes showing non-universality would indicate physics beyond the SM. This universality can be tested in
semileptonic $B$ meson decays involving a $\tau$ lepton, through
the ratio of branching fractions
\begin{equation}
R(D^{(\ast)}) \equiv  \frac{{\cal{B}}( B \to D^{(\ast)} \tau \nu_\tau)}{{\cal{B}}( B \to D^{(\ast)} \ell  \nu_\ell)}
\label{ratiotau0}
\end{equation}
The denominator is the average for $\ell \in \{e, \mu\}$. This ratio
 is typically used instead of the absolute branching fraction
of $ B \to D^{(\ast)} \tau  \nu_\tau$ decays, in order to cancel  uncertainties common to the numerator and the denominator.
These include the CKM matrix element and several theoretical uncertainties on hadronic form factors and experimental reconstruction effects.

In  the SM, values for $R(D^\ast)$ have been calculated already in 2012 by means of HQE, giving \cite{Fajfer:2012vx} \beq R(D^{\ast}) = 0.252 \pm 0.003 \eeq A later lattice estimate yields \cite{Na:2015kha} \beq R(D^{\ast}) = 0.300 \pm 0.008 \eeq
The discussion on different parameterizations, briefly outlined in Sect. \ref{subsectionExclusive decays}, has prompted in 2017 new SM determinations in the BGL parameterization \cite{Bigi:2017jbd, Jaiswal:2017rve, Bernlochner:2017jka}. These are generally consistent with the old predictions for $R(D^\ast)$.
Their arithmetic average, as given by HFLAV \cite{HFLAV2018} is
\begin{equation}
R(D^\ast) = 0.258 \pm 0.005
\end{equation}

As far as $R(D)$ is concerned, lattice  SM predictions \cite{Lattice:2015rga, Na:2015kha} have been averaged by the FLAG collaboration \cite{Aoki:2016frl}, yielding
\begin{equation}
R(D) = 0.2300 \pm 0.008
\end{equation}
As before, recent calculations have been performed in the BGL parameterization \cite{Bigi:2016mdz,Jaiswal:2017rve, Bernlochner:2017jka}, and their arithmetic HFLAV average \cite{HFLAV2018} is
\begin{equation}
R(D) = 0.299 \pm 0.003
\end{equation}

 Exclusive semi-tauonic $B$ decays were
first observed by the Belle Collaboration in 2007 \cite{Matyja:2007kt}.
Subsequent
analysis by Babar and Belle \cite{Aubert:2007dsa, Bozek:2010xy,Huschle:2015rga} measured
 branching fractions above, although consistent with, the SM predictions.
 In 2012-2013
Babar
 has measured
$R(D^{(\ast)})$ by using  its full data sample \cite{Lees:2012xj, Lees:2013uzd},
and reported a significant excess over the SM expectation, confirmed in 2016
 by the first measurement of $R(D^\ast)$
using the semileptonic tagging method (Belle \cite{Sato:2016svk}).
In 2015 a confirmation came also by the  LHCb collaboration, who has studied the decay $\bar B \to D^{\ast +} \tau \bar \nu_\tau$ with $ D^{\ast +} \to  D^{0}\pi^+$ and $\tau \to \mu \nu_\tau \bar \nu_\mu$ in $pp$ collisions
\cite{Aaij:2015yra}.

In 2016, the Belle collaboration  reported
 a new measurement
 in the hadronic
$\tau$
decay modes \cite{Hirose:2016wfn},
 statistically independent of the previous Belle
measurements,  with  a  different  background  composition.
These results
are consistent with the theoretical predictions of the SM in Ref. \cite{Fajfer:2012vx}
 within 0.6$\sigma$ standard deviations.
They also reported the first  measurement of the
$\tau$ lepton polarization in the decay $\bar B \to D^\ast \tau^- \bar \nu$  \cite{Hirose:2016wfn}, which is again compatible with SM expectations \cite{Tanaka:2012nw}.
Last year, the LHCB collaboration has measured $R(D^{\ast -})$
 in agreement with the SM
prediction \cite{Aaij:2017uff, Aaij:2017deq}.

By averaging the most recent measurements  \cite{Lees:2012xj, Lees:2013uzd,Huschle:2015rga,Aaij:2015yra, Sato:2016svk, Hirose:2016wfn, Aaij:2017uff, Aaij:2017deq},   the HFLAV Collaboration has found \cite{HFLAV2018}
\bea
R(D^\ast) &=& 0.306   \pm 0.013 \pm 0.007 \\
R(D)  &=& 0.407 \pm 0.039 \pm 0.024  \qquad \qquad
\label{ratiotau}
\eea
where the first uncertainty is statistical and the second one is
systematic. $R_D$ and $R_{D^\ast}$  exceed the SM
 values
  by about  2$\sigma$ and 3$\sigma$, respectively.
If one consider both deviations, the tension rises to about 4$\sigma$.
At Belle II a better understanding of
backgrounds tails under the signal  and a reduction of the uncertainty  to 3\% for $R_{D^\ast}$  and 5\% for   $R_D$ is expected at 5 ab$^{-1}$ \cite{DeNardo:2018glk}.

\section*{Acknowledgements}

 This work received partial financial support  from MIUR under Project No. 2015P5SBHT and from the INFN research initiative ENP.

\bibliographystyle{JHEP}
\bibliography{VxbRef}

%



\end{document}